# Exploring the spatial, temporal, and vertical distribution of methane in Pluto's atmosphere

E. Lellouch[1], C. de Bergh[1], B. Sicardy[1], F. Forget[2], M. Vangvichith[2], H.-U. Käufl[3]

[1]Laboratoire d'Études Spatiales et d'Instrumentation en Astrophysique (LESIA), Observatoire de Paris, CNRS, UPMC, Université Paris Diderot, F-92195 Meudon, France ; emmanuel.lellouch@obspm.fr

[2] Laboratoire de Météorologie Dynamique, Université Paris-6, 4 place Jussieu, F-75252 Paris Cedex 05

[3]European Southern Observatory (ESO), Karl-Schwarzschildst. 2, D-85748 Garching, Germany

## Abstract

High-resolution spectra of Pluto in the 1.66 µm region, recorded with the VLT/CRIRES instrument in 2008 (2 spectra) and 2012 (5 spectra), are analyzed to constrain the spatial and vertical distribution of methane in Pluto's atmosphere and to search for mid-term (4 year) variability. A sensitivity study to model assumptions (temperature structure, surface pressure, Pluto's radius) is performed. Results indicate that (i) no variation of the $CH_4$ atmospheric content (column-density or mixing ratio) with Pluto rotational phase is present in excess of 20% (ii) $CH_4$ column densities show at most marginal variations between 2008 and 2012, with a best guess estimate of a ~20 % decrease over this time frame. As stellar occultations indicate that Pluto's surface pressure has continued to increase over this period, this implies a concomitant decrease of the methane mixing ratio (iii) the data do not show evidence for an altitude-varying methane distribution; in particular, they imply a roughly uniform mixing ratio in at least the first 22-27 km of the atmosphere, and high concentrations of low-temperature methane near the surface can be ruled out. Our results are also best consistent with a relatively large (> 1180 km) Pluto radius. Comparison with predictions from a recently developed global climate model GCM indicates that these features are best explained if the source of methane occurs in regional-scale $CH_4$ ice deposits, including both low latitudes and high Northern latitudes, evidence for which is present from the rotational and secular evolution of the near-IR features due to $CH_4$ ice. Our "best guess" predictions for the *New Horizons* encounter in 2015 are: a 1184 km radius, a 17 µbar surface pressure, and a 0.44 % $CH_4$ mixing ratio with negligible longitudinal variations.

1. Introduction

Despite being much colder and ~300 times thinner, Pluto's atmosphere shares key characteristics with Mars'. Fundamentally, both are "equilibrium atmospheres", in which the dominant – and most volatile – icy component, i.e. $CO_2$ on Mars and $N_2$ on Pluto, is in vapor pressure equilibrium with the atmosphere. On both objects (and on Triton as well), the ice temperature is defined by the balance between insolation, thermal radiation, thermal conduction, and latent heat exchanges (sublimation or condensation) between atmosphere and surface. These atmospheres are "thick", in the sense that latent heat transport between summer and winter hemispheres is efficient enough to maintain weak lateral pressure gradients and surface frosts at a constant temperature over the globe. In that respect, Mars', Pluto's, and Triton's atmospheres stand in sharp contrast with Io's "thin" (1-10 nbar maximum) $SO_2$ atmosphere, which is also (at least partly) produced by sublimation of surface $SO_2$ but where latent heat terms are negligible in the energy budget; this leads to spatial variations of the frost temperature by tens of degrees and of the atmospheric pressure by orders of magnitude (Ingersoll 1990). The ice/atmosphere heat balance being modulated by orbit obliquity and/or eccentricity, all these sublimation-driven atmospheres exhibit seasonal variations of their surface pressure. While fully documented for Mars (Mars "pressure cycle"), these variations are much less characterized but still observationally established for the other objects, in particular at Pluto, for which a steady pressure increase over the last 25 years has been unambiguously detected (Sicardy et al. 2003, Elliot et al. 2003, Young et al. 2008, Person et al. 2008, Young 2013, Olkin et al. 2013). Similarly, at Triton, evidence for a factor-of-2 pressure increase over 1989-1997 has been found (Olkin et al. 1997, Elliot et al. 2000). Even with this limited time coverage compared to the orbital period, seasonal variations of the surface pressure appear to be, as expected, much larger on Pluto/Triton than on Mars, where the temperature of the $CO_2$ polar caps is approximately determined by its annually-averaged insolation and therefore undergoes little variations (Ingersoll 1990). Seasonal evolution on Mars also manifests itself in a variety of polar processes, with the formation of dark spots, fans and blotches during the retreat of the south-polar seasonal $CO_2$ ice cap (see e.g. Pilorget et al. 2011). Changes in the visual appearance of Pluto are of course much less documented, but temporal changes in Pluto's light curves and especially B-V color, not amenable to viewing geometry effects, provide strong evidence for real changes expected to result from volatile transport (Buie et al. 2010a, 2010b).

A second key common feature between Mars and Pluto (and again, Triton) is the additional presence of secondary, less volatile, surface species ($H_2O$ on Mars and $CH_4$/CO on Pluto and Triton), which are also sources of minor gases for the atmosphere through sublimation. Orbital effects drive variations with time and latitude of the temperature of these secondary ices, resulting in cyclic variations of their atmospheric content. Due to their much lower volatility and atmospheric abundance, these minor species do not buffer the corresponding surface ices as the major gases ($CO_2$ or $N_2$) do. Therefore, spatial/temporal variability of these ices temperature may be larger, although this depends also on whether they are intimately mixed with or thermally/geographically separated from the major ices. Mars water cycle is well characterized and understood (including the role of water clouds in the hemispheric transport of water, e.g. Montmessin et al. 2004). Much less is known for Triton and Pluto, though Lellouch et al. (2010) measured spectroscopically that at Triton the $CH_4$ partial pressure in 2009 was ~ 4 higher than determined by Voyager 20 years earlier, which they interpreted as due to the warming of $CH_4$-rich icy grains as Triton passed southern summer solstice in 2000.

Methane was the first volatile species discovered on Pluto as a surface species (Cruikshank et al. 1976) but its identification as a minor species had to await the discovery of $N_2$ ice on the surface in much larger amounts (Owen et al. 1993). Since then, detailed studies of the surface visible and near-IR spectra revealed the complexity of the methane surface distribution, exhibiting large lateral and vertical variability (Douté et al. 1999, Olkin et al. 2007, Merlin et al. 2010) as well as temporal evolution (Grundy and Buie 2001, Grundy et al. 2013). Essentially, methane exists on Pluto's surface in "pure" or "diluted" form[1], and in regions where the diluted form prevails, a large vertical gradient of the methane concentration in $N_2$ occurs, enriching the uppermost surface layers in methane. This complex situation has implications on the surface control of the methane atmospheric content. Regions where methane is in "pure" form, usually associated with relatively dark areas on Pluto images, are expected to be warmer (on the dayside) than the isothermal $N_2$ ice, not only because of their lower albedo, but also because the effect of sublimation cooling is minor (Stansberry et al. 1996); they therefore serve as preferential sources of atmospheric methane. Furthermore, the methane distribution may also affect the $N_2$ pressure cycle, as it has been proposed that the thin methane-rich surface layer could inhibit the sublimation of the underlying $N_2$ frost (Spencer et al. 1997, Trafton et al. 1997).

The first detection of atmospheric methane was reported from 1.66 μm spectroscopy ($2\nu_3$ band) with CSHELL/IRTF (Young et al. 1997), leading to a rough estimate of the $CH_4$ column density (1.2 cm-am within a factor of 3–4, assuming a mean gas temperature of 100 K). Improved observations of the same band obtained with CRIRES/VLT on August 1, 2008, including the detection of numerous individual methane lines, up to J = 8, made it possible to separate temperature and abundance effects (Lellouch et al. 2009, hereafter Paper I). Furthermore, the combined analysis of these data with stellar occultation curves indicated a $CH_4/N_2$ mixing ratio $q_{CH4}$=0.5±0.1 %. Similar numbers were obtained from additional (but of slightly lower quality) data taken on August 16, 2008. A consistent but less accurate $q_{CH4}$ = 0.6 (+0.6/-0.3) % mixing ratio was also obtained by Lellouch et al. (2011a) from observations at 2.3 μm. Even before its detection, methane had been recognized as the key heating agent in Pluto's atmosphere, able to produce a warm upper atmosphere (~100-120 K) and a sharp thermal inversion (Yelle & Lunine 1989, Lellouch 1994, Strobel et al. 1996). Building on these physically-based models, Zalucha et al. (2011a, 2011b) showed that an average $q_{CH4}$ ~ 0.5 % is indeed adequate to explain the atmosphere structure indicated from stellar occultation data, without having to invoke haze opacity. As a matter of fact, although some of the early interpretations of the occultation curves (e.g. Elliot and Young 1992) called for haze absorption, a convincing case for haze was obtained only from one dataset. The 2002 Aug. 21 occultation was observed from several telescopes atop Maunea Kea, from which B- to K-band light curves gave credible evidence for a wavelength-dependent absorber (Elliot et al. 2003). Subsequent occultations, however, failed to confirm this result; in particular the observation of a non-chromatic central flash in the July 31, 2007 occultation argued against significant haze contribution (Olkin et al. 2007)

Given the complexity of the methane surface distribution, and in particular the presence of isolated methane ice "patches" that can serve as sources for atmospheric methane, Pluto's atmospheric methane could conceivably exhibit spatial (latitudinal and longitudinal) variations. Another possibility is that of a vertical non-uniformity of the methane mixing profile. This is observed for Mars atmospheric water, which shows a general decrease of its mixing ratio with altitude in relation with the decreasing temperatures in Mars' troposphere, although it may be rather uniformly mixed near the warm perihelion and occasionally exhibit a complex (e.g. with detached layers) or supersaturated distribution (Maltagliati et al. 2013) . Albeit not well known, Pluto's thermal structure is very different from Mars', exhibiting at most a shallow troposphere (Paper I) and an extended stratosphere primarily resulting from the heating/cooling properties of methane. If the stratosphere is a region of weak mixing, it could plausibly be depleted in methane.

[1] This simplified terminology describes the fact that $N_2$ and $CH_4$ may form solid solutions that can be dominated by either of them (Douté et al. 1999).

Following our VLT/CRIRES 2008 observations presented in Paper I, we acquired additional spectra in 2012 at different longitudes with the following primary goals: (i) search for rotational variation of the methane abundance (ii) search for temporal variability on a mid-term (4 year). Additionally our data put some constraints on the extent to which methane could depart from uniform mixing.

## 2. Observations and data reduction

Repeated observations of Pluto were obtained in 2012 in service mode with the cryogenic high-resolution infrared echelle spectrograph (CRIRES, Käufl et al. 2004) installed on the ESO VLT (European Southern Observatory Very Large Telescope) UT1 (Antu) 8.2 m telescope. CRIRES was used in adaptive optics mode (MACAO) with a 0.4" spectrometer slit. The instrument spectral resolution is approximately given by R = 96,000 x 0.2 arcsec / D, where D is the seeing-dependent angular diameter of the source after AO correction. In our case, this typically provided R ~60,000.

As in our 2008 observations, we focused on the $2\nu_3$ band of methane near 1.66 µm. The instrument consists of four Aladdin III InSb arrays, covering approximately 8 nm each. However, unlike in 2008, we used a fast ("windowed") readout mode in which only detectors 2 and 3 are read (and windowed). This mode had resulted in a noticeable gain in sensitivity during our observations of the 2.3µm range in 2010 which led to evidence for CO in Pluto's atmosphere (Lellouch et al. 2011a). However, for equal on-source time as in 2008 (70 mn per Pluto visit), it turned out that the 1.66 µm observations from 2012 did not have better S/N (rather, slightly worse) than in 2008. The spectral range of the 2012 observations was 1642–1650 and 1652–1659 nm. Although this resulted in a much smaller number of lines than in 2008 (in particular the Q-branch was not re-observed), the spectral interval includes the R(1), R(2), R(3), R(5), R(6) and R(7) lines, sampling a broad range in lower energy level (up to 293 $cm^{-1}$, equivalent to 420 K) as well as a number of weaker lines of low energy belonging to other bands.

Initially, our program included 6 visits to Pluto (Runs A-F), that were supposed to be scheduled (in service mode) to sample evenly all Pluto rotational longitudes[2]. For that we had provided 6 lists of dates (corresponding to 60° bins in Pluto longitudes) with the instruction that observations should be acquired on one date per list. The list of dates was also built to ensure a large (>15 km/s) topocentric velocity, permitting a proper separation of the Pluto methane lines from their telluric counterparts. For the most part, our program was appropriately executed, except that (i) two of the longitude bins (L = 120-180 and 180-240) were not observed (ii) one (L = 240-300) was observed twice and at very similar longitudes. Overall, this resulted in five spectra, with longitudinal redundancy for two of them (spectra D

[2] We use the orbital convention of Buie et al. (1997) in which north is defined by the direction of the system angular momentum vector. Therefore the North Pole is currently facing the Sun, and longitudes, noted L, are East longitudes

and E). Table 1 summarizes the observing conditions (UT date, longitude, Pluto air-mass, topocentric and heliocentric velocity, water vapor content) for the 5 spectra, and recalls the same parameters for the two observations from 2008 (labeled "1" and "16" after their observing date in August 2008). Note that these two observations were performed in "normal" (non-windowed) mode and thus covered four spectral ranges at 1642–1650, 1652–1659, 1662–1670, and 1672–1680 nm (see Paper I).

Table 1: Observational parameters

| Run | UT date (yy/mo/day – begin/end UT) | Long. L | Air mass | $V_{topo}$ (km/s) | $V_{helio}$ (km/s) | $H_2O$ (ppt-mm) |
|---|---|---|---|---|---|---|
| A | 2012/08/14 01h35-02h55 | 18 | 1.02 | 21.692 | 0.982 | 1.5 |
| B | 2012/04/07 08h30-09h50 | 72 | 1.04 | -28.391 | 0.982 | 8 |
| D | 2012/08/10 01h25-02h45 | 244 | 1.01 | 20.130 | 0.962 | 2.2 |
| E | 2012/09/10-11 23h35-0h55 | 245 | 1.02 | 29.025 | 0.965 | 2.1 |
| F | 2012/08/15 01h48-03h08 | 322 | 1.04 | 22.062 | 0.967 | 1.6 |
| 1 | 2008/08/01 03h10-04h30 | 301 | 1.15 | 19.999 | 0.844 | 2 |
| 16 | 2008/08/16 0h55-02h20 | 181 | 1.03 | 24.895 | 0.862 | 2 |

At all dates, a comparison spectrum on a A3V star (HR 5917, H mag = 5.283) was recorded at a similar air-mass as Pluto in order to validate our telluric atmospheric model, and precisely calibrate the wavelength scale. Its spectral resolution was determined from the width of the telluric lines. Resolving powers of 60000, 65000, 70000, 65000, 65000, were found for the spectra of runs A, B, D, E and F, respectively, indicating effective source sizes of 0.27"-0.32". The same values were applied to the associated Pluto spectra; this is valid because Pluto's finite size (~0.1") implies a negligible additional smearing of this source size.

All data were reduced using the standard steps of the CRIRES pipeline suited to long-slit spectroscopy, including corrections for darks, flat-field, image recombination, replacement of bad pixels and outliers, spectral extraction of the cleaned images, and wavelength calibration. Along with the 5 spectra from 2012, we re-reduced the 2008 data in a consistent manner, leading to cosmetic improvements compared to the original data reduction in Paper I. Fig. 1 shows all seven spectra – uncorrected for telluric and solar lines – over 1642-1649 and 1653-1660 nm in the Pluto velocity frame. Each spectrum clearly shows 6 lines from the $2\nu_3$ band (R(1) – R(3) and R(5) – R(7)) flanked by their telluric counterparts (on the blue side, given the values of the topocentric velocities, except for spectrum B). Additional weak methane features are present, as discussed later. Based on simple inspection of line depths, no obvious variability from spectrum to spectrum is present.

## 3. Model

Spectra were modeled using a standard radiative transfer code. Radiation was integrated over angles, using the classical formulation in which the two-way transmittance is expressed as $2E_3(2\tau)$, where $\tau$ is the zenithal optical depth of the atmosphere. This approach (e.g. Young et al. 2001) implicitly assumes negligible scattering and no limb-darkening of the solar component. The Pluto model was then monochromatically multiplied by a model of the telluric transmission and of the solar spectrum, taking into account the appropriate topocentric and heliocentric Doppler shifts. The solar spectrum was taken from Fiorenza and Formisano (2005), and the telluric spectrum was calculated by using the LBLRTM code (Clough et al. 2005; http://rtweb.aer.com/lblrtm.html) at the relevant Pluto air mass and with the appropriate water vapor amount (Table 1). Overall the model was identical to that of Lellouch et al. (2009) except that we used updated versions of LBLRTM and of the methane low-temperature molecular list. Specifically we used the “80 K” line list from Campargue et al. (2013), extrapolating the intensities from 80 K to the relevant Pluto atmosphere temperatures by making use of the given “empirical” lower energy values. In most of our models, methane was assumed to be uniformly mixed in the whole atmosphere, but we also more briefly studied alternate models. The resulting synthetic spectra including the Pluto, telluric and solar lines were convolved to the instrumental resolution for directly comparison to the data.

Except for the established presence of a strong inversion layer joining Pluto’s surface or low-altitude troposphere, Pluto’s atmosphere thermal structure is not well known. In particular, the surface pressure at a given epoch and the lower atmosphere structure, including the possible existence of a troposphere, remain uncertain. In Paper I, the multiplicity of methane lines made it possible to determine a mean gas temperature (T) for methane, and in the framework of an isothermal model, an associated $CH_4$ column density ($a_{CH4}$). However, and although the temperature determination was relatively accurate (90+25/-18 K), the associated column density was uncertain by a factor ~1.7 in both directions ($a_{CH4}$ = 0.75+0.55/−0.30 cm-am). This stems from the fact that methane lines are extremely narrow (Doppler-shaped) and often saturated, meaning that a large increase (resp. decrease) in column density is required to compensate for a small decrease (resp. increase) of temperature in producing the same line depths. A second step in Paper I was to combine spectroscopy with constraints from stellar occultations. For that synthetic occultation curves were modeled using a variety of thermal profiles, and their general features (e.g. residual flux level, “knee” associated with the thermal inversion, caustic spikes, central flash) were examined in the light of existing occultation data, considering in particular the high quality 21 Aug. 2002 CFHT and 12 June 2006 AAT light curves. The main results of this exercise were (i) to constrain families of acceptable thermal profiles and the surface pressure to the 6.5-24 μbar range in 2008 (ii) to show that Pluto’s troposphere, if existing, is shallow at best (< 17 km) (iii) to determine the methane mixing ratio to be $q_{CH4}$ =0.5±0.1 %. The fact that the mixing ratio could be better determined than the column density is due to the following. The surface pressure is uncertain by a factor ~2 in both directions. Deeper models have mean temperatures colder than shallower models; therefore they require larger methane columns to fit the spectrum, which partly compensate for the higher surface pressure, reducing the uncertainty in the mixing ratio. Atmospheres deeper than 24 μbar in 2008 (associated with Pluto radii less than 1169 - 1172 km) were rejected because their mean temperature is too cold to be consistent with the methane spectrum.

Here we used a somewhat different approach. As the new data include fewer lines than in the original observation (although almost the same range of lower energy levels, up to J = 7, is covered) and also have a somewhat lower S/N, we do not repeat the isothermal fits. Indeed, based on the results from Paper I we can anticipate that this approach would lead to relatively large (factor ~2 or more) uncertainties in the retrieved methane columns. As the goal of our observations is to search for rotational and temporal variability of the methane abundance, a better approach is to assume some temperature structure, and use it to fit all spectra. There are several reasons to believe that lateral variability of the temperatures is minimal. Except for possible but unknown topographic effects, the surface pressure must be spatially uniform, as discussed above. When available, multi-site observations of a given occultation reveal no large variations (>10 K) or significant trends of the upper atmosphere temperature with latitude or morning/evening, despite the fact that the ingress/egress points may occur at places with highly variable instantaneous insolation (Millis et al. 1993, Young et al. 2008). This supports the fact that the dynamical redistribution of solar heat erases most of lateral temperature contrasts – except at the surface itself where the darkest regions may be up to ~25 K warmer than the $N_2$ ice (Lellouch et al. 2000). This is supported by General Circulation Models (Vangvichith and Forget 2013) which find latitudinal variations of the zonal mean temperatures at the 1-2 K level only, and small amplitudes (< 0.5 K) of thermal tides. Although the thermal structure itself is poorly known, we can expect that uncertainties will cancel out when studying possible variations of the methane content. However, as the surface pressure is known to vary with time, it seems worthwhile to consider different atmospheric models for 2008 and 2012.

The considered thermal structures are shown in Fig. 2, in temperature-radius, temperature-pressure and pressure-radius forms. Two different thermal profiles were used, meant to represent possible thermal states of the atmosphere in 2008 and 2012. For 2008, we utilized the same thermal profile as adopted in Lellouch et al. (2011a) for their analysis of the 2.3 μm spectrum. This profile (of course one of the solution profiles in Paper I, see their Fig. 4) has a stratospheric gradient of 6 K/km and a wet tropospheric adiabat (gradient -0.1 K/km).

To represent the 2012 atmosphere we used a thermal profile constrained by stellar occultations observed on July 18, 2012 and May 4, 2013 (Dias de Oliveira et al. 2013, Sicardy et al. in prep.). Very high S/N observations with NACO/VLT of the July 18, 2012 event provide a temperature "template" in the form of a T(r) curve, but with uncertainties on the absolute vertical scale (i.e. r) of several kilometers due to uncertainties in the impact parameter (the distance of closest approach of the observing station to Pluto's shadow center). A noteworthy feature of the temperature profile is the presence of a mesosphere with temperatures progressively decreasing above a 110 K stratopause. The May 4, 2013 event provides data with lower S/N compared to 2012, while being fully compatible with the 2012 T(r) template. Moreover, as it was recorded from 6 sites that covered both northern and southern offsets relative to Pluto's shadow center, it provides a geometric reconstruction of the event with an accuracy of about ±1 km in the altitude scale. The resulting temperature-pressure-density profile extends down to 1190 km, where the pressure is 11.3±0.07 μbar, the temperature is 60 K, and the thermal gradient is 5.5 K/km. For our purpose, the profile was extrapolated below this altitude in a way similar to the "2008" profile (see Paper I).

Pluto's radius and surface pressure are unknown, so in addition to the two temperature profiles, we considered two values of the radius, 1175 km and 1188 km. The first value is close to the minimum acceptable Pluto radius associated to the "2008" temperature profile (1172 km), according to Paper I. The second value implies a surface pressure of 9.4 μbar, close to the minimum pressure for 2008 (7.3 μbar). Moreover, with the "2012 profile", the extrapolated temperature at 1188 km reaches 50 K, close to the mean dayside Pluto surface temperature (Lellouch et al. 2000); therefore 1188 km is probably near the maximum value of Pluto radius. Therefore, the two assumptions for Pluto radius span rather well the range of possibilities, and as shall be seen, a radius as small as 1175 km can be excluded by our new data.

Pressures at 1275 km, 1188 and 1175 km are 1.4, 9.4 and 19.5 μbar, respectively, in the 2008 profile, and 2.3, 12.3 and 25 μbar, respectively, for 2012 (i.e. a level-dependent factor of ~1.6-1.3 change). This is in line with the continuing pressure increase in Pluto's atmosphere (Olkin et al. 2013). From their own observations of the May 4, 2013 occultation, Olkin et al. infer a 2.7±0.2 μbar pressure at 1275 km (and an upper atmosphere T = 113 ± 2 K, consistent with ours), and find a pressure change by a factor ~1.3 over 2007-2013 at this level ($p$ = 2.1 μbar in July 2007). Note finally that in the entire stratosphere, the “2012” profile is warmer than the “2008” profile by about 15 K. While there remains some uncertainty on the magnitude of the pressure evolution, our results on the methane abundances can be easily transposed to other assumptions on the total pressure, as explained below.

## 4. Results

### 4.1. Sensitivity to atmospheric model

To first assess the effect of the uncertain thermal structure on the methane abundance indicated by the spectra, we first used the two temperature profiles (along with the two values of the surface pressure, i.e. 4 cases in total) to model the *average* of the five spectra recorded in 2012. For this, the five spectra were aligned in the Pluto velocity frame (see Fig. 1) and averaged. Because each spectrum has a different topocentric velocity, the outcome of this procedure is to increase the S/N on the Pluto methane lines and to smear the telluric and solar lines. The same exact process was applied to the model and the model-observation comparison yielded the best fit methane mixing ratio ($q_{CH4,}$ assumed vertically uniform) and associated column density. Results are given in Table 2 (which also includes results from the same analysis of the spectrum of August 1, 2008). They are illustrated in Fig. 3 which shows fits in the range of the R5-R7 lines, which are the most sensitive to the methane abundance.

| Data set | Atmospheric model | Radius (km) | $CH_4$ mixing ratio ($q_{CH4,}$ %) | $CH_4$ column density (cm-am) |
|---|---|---|---|---|
| 2012 (A-F) | 2012 | 1188 | 0.32 ± 0.04 | 0.57 ± 0.08 |
| | | 1175 | 0.32 ± 0.04 | 1.09 ± 0.15 |
| | 2008 | 1188 | 0.49 ± 0.06 | 0.68 ± 0.08 |
| | | 1175 | 0.49 ± 0.06 | 1.33 ± 0.16 |
| 2008 Aug. 1 | 2012 | 1188 | 0.41 ± 0.06 | 0.73 ± 0.11 |
| | | 1175 | 0.39 ± 0.06 | 1.32 ± 0.19 |
| | 2008 | 1188 | 0.60 ± 0.10 | 0.84 ± 0.14 |
| | | 1175 | 0.60 ± 0.10 | 1.63 ± 0.27 |

Table 2: Fits of the 2012 average spectrum and of the 2008 Aug. 1 spectrum as a function of the atmospheric model and surface radius. See text for explanations.

The following conclusions can be drawn from inspection of Table 2 and Figure 3. First, for a given temperature profile, the inferred $CH_4$ mixing ratio is largely *independent on the radius* at which the atmosphere joins with the surface. (In contrast the resulting methane column increases with decreasing surface radius – by almost a factor of 2 between the two cases considered – confirming the statement in Paper I that the $CH_4$ mixing ratio is better determined than its column density). Essentially this is because any atmosphere “added” below a certain pressure level is “invisible” in the strong methane lines of the R branch which already show saturated absorption. However this behavior is no longer true in regions of weak methane absorption. This occurs in particular in a series of weak lines of lower energy levels that are present in-between the J-manifolds of the 2 $\nu_3$ band. When the atmosphere is “deep”, the modeled absorption in these weak lines continues to increase and at some point becomes inconsistent with the measurement. This effect, which could already be seen in the data of Paper I (see the Doppler-shifted absorption at 1643.43 nm in Fig. 3 of that paper), is demonstrated by comparing the first row in Fig. 3 (surface radius R = 1188 km) to the second row (R = 1175 km): for a given abundance, models for the two cases are indistinguishable in the R5, R6 and R7 lines; however models with R = 1175 km produce too much absorption near 1643.32, 1645.77, and 1646.15 nm. Based on this, we would favor "shallow" atmospheres and conclude that Pluto's radius is not smaller than ~1180 km. As we will see later, the same kind of argument can be used to constrain departures from uniformity of the methane mixing profile.

A second conclusion from Table 2 is that *for a given surface radius*, the retrieved methane column density is only mildly sensitive to the adopted thermal structure, being about 20 % lower for the warmer 2012 profile compared to the 2008 profile. This difference is marginally significant given the error bars on the derived columns, which are also of order 20 %. The retrieved mixing ratios are more strongly dependent upon the assumed atmospheric model, and the difference is predominantly an effect of the different surface pressures for the two models. Indeed, the mixing ratios for the case of the "2008 thermal profile" are systematically higher than those found for the "2012 thermal profile" by a factor ~1.5. Out of this 50 % effect, 30 % is due to the ratio of the surface pressures, and 20 % come from the difference in columns discussed above.

### 4.2. Search for rotational and temporal variations of atmospheric methane

Table 3 presents the methane column density and mixing ratio $q_{CH4}$ derived from the 7 individual observations, assuming a Pluto radius R = 1188 km. Observations for each year are fit with the adopted thermal profile for the corresponding year. Based on the above considerations, the value of the radius has no impact on the retrieved $q_{CH4}$, while for another surface radius R', the $CH_4$ columns given in Table 3 should be rescaled by $p$(R') / $p$(R), where $p$ the is surface pressure. Conversely, the mixing ratios in Table 3 hold for the pressure distribution in the 2008 and 2012 profiles (having $p$(1188 km) = 9.4 μbar in 2008 and $p$(1188 km) = 12.3 μbar in 2012), and could be rescaled as 1/ $p$ for any other adopted values of the surface pressure.

| Observing Run | Atmospheric model (profile / radius) | $CH_4$ mixing ratio ($q_{CH4,}$ %) | $CH_4$ column density (cm-am) |
|---|---|---|---|
| 2012 – Run A | 2012 / 1188 km | 0.37 ± 0.07 | 0.67 ± 0.13 |
| 2012 – Run B | 2012 / 1188 km | 0.29 ± 0.05 | 0.51 ± 0.08 |
| 2012 – Run D | 2012 / 1188 km | 0.29 ± 0.05 | 0.51 ± 0.08 |
| 2012 – Run E | 2012 / 1188 km | 0.42 ± 0.07 | 0.75 ± 0.13 |
| 2012 – Run F | 2012 / 1188 km | 0.39 ± 0.06 | 0.70 ± 0.11 |
| 2008 – Run 1 | 2008 / 1188 km | 0.60 ± 0.10 | 0.84 ± 0.14 |
| 2008 – Run 16 | 2008 / 1188 km | 0.50 ± 0.12 | 0.70 ± 0.16 |

Table 3: Fit solutions of the 7 individual spectra. See text for explanations.

Results from Table 3 are shown graphically in Fig. 4. They show first that little, if any at all, changes have occurred in the methane content from 2008 to 2012. Table 3 indicates mean column densities of 0.77±0.09 cm-am in 2008 (averaged over the two measurements) and 0.57±0.08 cm-am in 2012 (averaged over five), i.e. a 1.35+0.40/-0.30 ratio. The evidence for change is therefore marginal, considering also that (i) the use of two different temperature profiles for the two epochs contributes to about half of the effect and (ii) the variation is also smaller if only data at similar longitudes are considered (see Fig. 4). Our best guess for the evolution of Pluto's global methane content is a ~20 % decrease over 2008-2012. This general constancy is also consistent with the fact that based on data recorded in July 2010 at 2.3 μm, and analyzed with the "2008" thermal profile, Lellouch et al. (2010) inferred $q_{CH4}$ = 0.6 % (though with a factor of 2 uncertainty). Second, methane shows at most small variations with longitude. There may be a hint for slightly larger (30-40 %) methane columns over L = 240-300 longitudes than on the opposite hemisphere, but this behavior is essentially within error bars and does not fit one data point (Run D, L = 244). We more conservatively conclude that disk-averaged methane column densities are constant with longitude within ±20 %. The right panel of Fig. 4 shows the same information in terms of the methane mixing ratio. Immediately visible is the fact that $q_{CH4}$ has apparently decreased over 2008-2012 by a factor of 1.7+0.5/-0.4. Although the marginal decrease of the methane columns contributes, this mostly results from the increase in pressure with time, by a factor 1.3 at 1188 km and 1.6 at 1275 km in our adopted atmospheric models.

### 4.3 Constraints on the vertical distribution of methane

In all the preceding, we have assumed that methane is vertically uniformly mixed with nitrogen. The same assumption was made in Paper I, upon the argument that (i) the source of methane is at the surface and (ii) its equivalent temperature (~90 K) implies that a significant fraction of it is in the upper atmosphere. As can be seen from fitting quality, there is no obvious contradiction to this assumption. Still, it is useful to explore to what extent $CH_4$ could depart from a uniform distribution, especially because due to its restricted troposphere and extended stratosphere, Pluto's atmosphere may be poorly mixed. To do so, we considered a series of models with R = 1188 km, schematically describing the distribution of methane as constant up to some cut-off altitude $z_0$. For each value of $z_0$, the methane mixing ratio from the surface to $z_0$ was adjusted to match the depth of the strong lines of the $2\nu_3$ band and the impact on the rest of the spectrum was examined.

Results are shown in Fig. 5 for the case of the Aug. 1, 2008 spectrum, using the "2008" thermal profile, and focusing on three diagnostic spectral intervals. Fits using $z_0$ = 1225 km and 1210 km are shown and compared with the uniform mixing case ($z_0 = \infty$). This clearly shows that $z_0$ = 1225 km provides a fit as good as $z_0 = \infty$. In contrast $z_0$ = 1210 km is readily excluded, as the enhanced mixing ratio required to fit the strong lines (e.g. the R(1) line at 1659.6 nm) leads to spurious additional absorptions throughout the spectrum. The minimum value of $z_0$ still consistent with the data is ~1215 km. In essence, when methane is restricted to the near-surface, its average temperature is too cold to match the data. Hence, methane is well mixed in a layer at least 27 km thick. Data from 2012, if analyzed with the "2012" thermal profile, lead to a similar though slightly less constraining result, i.e. that $z_0$ > 1210 km. This comes from the fact that as noted above the 2012 thermal profile is warmer by ~20 K compared to 2008 near 1200 km. Therefore, an appropriate mean methane temperature can be achieved with methane extending to slightly less high altitudes in the warm part (>100 K) of the profile.

The above results are valid for the 1188 km radius case (which as mentioned above should be close to the maximum Pluto radius). Since the atmosphere below this level is cold (~40-50 K), assuming any smaller value of the radius in the 1175 – 1188 km range would require an even broader region of uniform mixing. As discussed previously in Sec. 4.1, a radius smaller than 1180 km is inconsistent with the spectra even for uniform mixing of the methane throughout the atmosphere.

The essential conclusion is that methane must be roughly uniformly mixed in a significant thickness of the atmosphere, at least 22-27 km and probably more. This rules out models when there is "much more" (in the sense more than uniform mixing would say) "cold" vs "warm" methane. At the 2013 Pluto science conference, Cook et al. (2013) argued for such a situation, specifically suggesting that their Keck/NIRSPEC observations indicated a total methane column of 1.8 x $10^{20}$ mol $cm^{-2}$, 94 % of which (i.e. 6.3 cm-am) at 40 K and 6 % (0.4 cm-am) at 100 K. Running explicitly this "2-layer" model readily shows that it is inconsistent with our data (see Fig. 5).

## 5. Discussion

As has been long recognized, a $CH_4$ / $N_2$ mixing ratio of several tenths of a percent is orders of magnitude greater than the ratio of their vapor pressures at any given temperature prevailing on Pluto, and in the framework of a surface-atmosphere equilibrium controlled by Raoult's law, the discrepancy is exacerbated by the fact that methane is a minor surface component. Based on the observation that "pure" and "diluted" methane co-exist on Pluto, two scenarios have been historically proposed to explain the elevated methane abundance (1) the existence of geographically separated patches of "pure and warm" methane, which are warmer than nitrogen-rich regions because less subject to sublimation cooling, thereby boosting the relative $CH_4$ / $N_2$ atmospheric content (Stansberry et al. 1996, Spencer et al. 1997) (2) the formation, in regions where methane is "diluted", of a thin uppermost surface layer (termed "detailed balancing") much enriched in methane, which inhibits the sublimation of the underlying, dominantly $N_2$, frost and leads to an atmosphere with the same composition as this frost (Trafton et al. 1997). Arguments in favor of the first scenario are that: (i) direct thermal measurements strongly suggest that methane ice can reach dayside temperature well in excess of 50 K (Lellouch et al. 2000, 2011b) and that (ii) simple calculations (Stansberry et al. 1996) suggest that pure methane patches covering only a few percent of Pluto's surface may be sufficient to maintain 0.1-1 % $CH_4$ mixing ratios in the atmosphere. Support for the second scenario may be found from the facts that (i) the atmospheric $q_{CH4}$ is indeed similar to the surface $CH_4$ / $N_2$ mixing ratio in regions of high methane ice dilution, and that (ii) these regions are indeed the place of a large positive gradient of the methane concentration towards the surface (Douté et al. 1999, Olkin et al. 2007).

Based on multi-night monitoring of Pluto's near-IR band spectrum and a study of its secular evolution, and building upon previous studies by Grundy and Fink (1996) and Grundy and Buie (2001), Grundy et al. (2013) established that the deeper methane near-IR bands show maximum absorption near L = 260 longitude, while shallower bands have maximum absorption progressively shifted further east (up to near L = 320 for the weakest bands; see sketch in Fig. 4). This suggests that larger particle sizes or greater $CH_4$ enrichments – needed to boost the depths of weaker $CH_4$ absorptions – occur preferentially on the Charon-facing hemisphere (L ~ 0). In contrast, the degree of methane dilution (as determined from band shifts) is minimum on the sub-Charon hemisphere and maximum on the anti-Charon hemisphere (L ~ 180), which is also where the $N_2$ and CO bands are strongest. Furthermore, Grundy et al. (2013) show that, on decadal timescales, Pluto's $CH_4$ ice absorption bands have deepened and the amplitude of their rotational variation has decreased. This is consistent with strong $CH_4$ absorption at high Northern latitudes progressively emerging into view as the sub-Earth latitude moves North (e.g. from ~9°N in 1992 to 48°N in 2012).

In the framework of the first scenario above, in which atmospheric methane is provided by sublimation of "pure" methane ice patches, one might naively expect larger (if anything) maximum methane gas columns at L = 260-320. This is not inconsistent with our observational results (Fig. 4), however, as mentioned before, we more conservatively conclude that disk-averaged methane column densities are constant with longitude within ±20%. This would mean that even if increased sublimation occurs at these longitudes, dynamical redistribution of methane occurs and largely erases longitudinal contrasts.

Pluto's methane cycle was recently investigated by means of a full-blown global climate model (GCM, Vangvichitch 2013, Forget and Vangvichitch 2013) inherited from the Martian LMD GCM (e.g. Forget et al. 1999). This model solves the 3-D hydrodynamical equations and includes a description and parameterization of all relevant physical phenomena, including radiative transfer (due to $CH_4$ and CO in Pluto's case), thermal molecular conduction, convection and turbulence, sub-surface conduction, $N_2$ and $CH_4$ surface sublimation/condensation exchanges, atmospheric condensation (in contrast, the thermodynamics of $N_2$ – $CH_4$ mixtures are not considered). Following previous studies (Grundy and Fink 1996, Lellouch et al. 2000, 2011b, Grundy and Buie 2001), Pluto's surface is depicted with three different "pure" units, composed respectively of $N_2$ ice, $CH_4$ ice, and tholins, and the geographical distribution retained by Lellouch et al. (2000) (also termed as "$g_2$" in Lellouch et al. (2011b) and shown here as background of Fig 4b) is adopted. In this distribution, methane ice is concentrated in two regions: the low-latitudes (within ±30°), with extended methane areas at L = 240-360 (in accordance with constraints from the near-IR), and the high Northern (> 50°N) latitudes, which serve as methane sources for the atmosphere. All three units are assumed to have a 0.9 bolometric emissivity, and their Bond bolometric albedos are taken from Lellouch et al. (2011b). Furthermore, the model is nominally run with a diurnal thermal inertia of 20 SI (again from that paper), and a seasonal thermal inertia of 500 SI, a bit smaller than but generally in line with the high thermal inertias invoked to account for the current pressure increase in Pluto's atmosphere (1000-3000 SI; Young 2013, Olkin et al. 2013). Therefore, in addition to the 3-D meteorological fields (temperature, winds…), the model is able to calculate the 3-D distribution of Pluto's methane and its evolution with time, in much the same way as the Martian GCM is able to successfully reproduce Mars' water cycle (Montmessin et al. 2004).

Results from the model (Forget and Vangvichitch 2013) compare generally well with our observational findings. The first (and spectacular) successful model prediction is that of a typical mixing ratio of several tenths of a percent; for example, for 2010, the vertically-averaged $q_{CH4}$ in the model is ~0.3 % at low latitudes, reaching 0.6 % at high Northern latitudes (since the surface pressure is expected to be spatially uniform, this implies that the methane column would vary by a factor of 2 from low to high Northern latitudes). This essentially means that the "warm patch" model is able to explain the elevated $CH_4$ atmospheric abundance and brings to light a nicely self-consistent picture: a distribution of methane ice deposits with optical and thermal properties consistent with a variety of constraints (visible imaging and light-curve, mutual events, visible and near-infrared spectroscopy, thermal light-curves) is able to sustain atmospheric $CH_4$ amounts in essential agreement with inferences from high-resolution spectroscopy and stellar occultation light-curves. Thus, at least to first order, more complex scenarios do not seem to be needed, such as the role of haze opacity in explaining the occultation data, or of the "detailed balancing" layer in controlling the methane gas abundance. In fact, as outlined above, except for the 2002 Aug. 21 occultation multi-color light curve measurements, convincing evidence for the haze has been lacking, although this does not rule out any role for haze/clouds for all times.

With the current orbital configuration (sub-solar latitude ~41° in 2008, 48° in 2012), Pluto's high northern latitudes are the place of maximum insolation, hence the existence of the high latitude methane patches is key to the atmospheric abundance. This has implications in two ways. First, as in the model, the high-latitude methane ice is rather uniform in longitude, the associated longitudinal variations of atmospheric methane are kept small, and further subdued by dynamical mixing. This is illustrated in Fig. 6a which shows the modeled variation of the mean apparent (as seen from Earth) atmospheric methane mixing ratio in 2010 as a function of longitude. Only a very weak maximum (+10 % compared to minimum) is apparent near L =0, in agreement with our lack of detection of longitudinal variability within ±20 %. Second, this near-polar reservoir plays an important role in controlling the secular evolution of $CH_4$ gases. This is shown in Fig. 6b where the apparent methane column density over 2000-2016 is calculated for the nominal model and compared to a model in which no $CH_4$ ice is assumed to be present northward of 60°N. In the nominal model, the $CH_4$ column declines from 2000 to reach a broad minimum of ~3 g $m^{-2}$ (i.e. 0.42 cm-am) over 2006-2012. This decline, due to the progressive cooling of the low-latitude $CH_4$ frosts at Pluto's, is then offset by the warming of the high-latitude $CH_4$ deposits as the sub-solar latitude progresses northwards. In the absence of high-latitude frosts, the decline of the $CH_4$ column with time persists, and $CH_4$ collapses by a factor > 5 over 2000-2015. Based on Fig. 6b, models predict that the methane column in mid-2008 was 10 % lower than in mid-2012 for the nominal case, and a factor 1.7 larger in the case where no $CH_4$ ice is present at high Northern latitudes. Given the error bars, our observational results are not entirely conclusive, but given the discussion in Section 4.2, we feel that they mostly indicate indistinguishable methane columns in 2012 vs 2008, in accordance with the nominal model. This is consistent with the observation, based on the secular variations of the $CH_4$ ice bands, that methane ice is present at high northern latitudes (Grundy et al. 2013). We note that the two scenarios predict diverging evolutions of the methane column over the upcoming years, so that a continuing monitoring of the gas methane should permit to validate our conclusions in a more definite way.

In spite of their large error bars, the initial observations of Young et al. (1997), performed in May 1992, may suggest further evidence that the methane column is time-variable. Analyzing their data in terms of a uniform layer at T = 100 K, the authors inferred a $1.20^{+3.15}_{-0.87}$ cm-am column density. Adapting our model to the condition of their data (including the proper topocentric / heliocentric velocities, Pluto air mass and spectral resolution) shows excellent consistency with their findings. Reanalyzing the data with T = 90 K (the mean methane temperature according to paper I), we find that these columns should be approximately multiplied by a factor 1.33, giving 0.44-5.8 cm-am, with a best fit value of 1.6 cm-am. While the large error bars prevent a definitive conclusion, this is suggestively larger than the 0.75±0.13 cm-am inferred from the same T = 90 K fit of the 2008 Aug. 1 data. This would confirm the GCM prediction that the $CH_4$ columns have initially decreased post-perihelion (Fig. 6a).

Regarding the $CH_4$ vertical profile, we find that the data are consistent with uniform mixing in a significant fraction of the atmosphere, at least 22-27 km and probably more, and rule out situations in which methane is concentrated in near-surface cold layers. This also agrees with the predictions of Forget and Vangvichith (2013) who find that, except for a ~5 km-thick methane-depleted near-surface layer, $CH_4$ is essentially uniformly mixed in the bulk of the atmosphere (at least from ~5 to ~40 km).

In combination with the continuing pressure increase of Pluto's atmosphere, the near-constancy of the $CH_4$ column density between 2008 and 2012 implies a decrease of its mixing ratio over the same period. Our adopted atmospheric models lead to a factor of $1.7^{+0.5}_{-0.4}$ decrease of the mixing ratio over 2008-2012, where most of the effect is due to the 1.3-1.6 times larger pressures in 2012. The factor of ~ 4 pressure increase since 1988 and our estimate above of the $CH_4$ column density in 1992 also give evidence for a general decrease of $q_{CH4}$ in the last 20 years. To first order, methane controls the stratosphere temperature structure. Therefore, the question rises whether a progressive decline of $q_{CH4}$ with time is still consistent with the observed "upper atmosphere" (i.e. stratopause) temperatures, which have never been observed to depart by more than ~10 K from a mean value of ~105 K (and with maximum temperatures of ~111-115 K in 2013). First, as originally demonstrated by Yelle and Lunine (1989), the thermostat role played by $CH_4$ (with absorption of solar light in the near-infrared bands balanced by cooling in the 7.7 μm $\nu_4$ band) is efficient enough that the upper atmosphere temperature is relatively insensitive to the methane mixing ratio. This is confirmed in the more detailed models of Strobel et al. (1996), Zalucha et al. (2011a, 2011b), and Zhu et al. (2014), where changes by a factor ~3 in $q_{CH4}$ lead to temperature changes well within 10 K. Interestingly, Zalucha et al. (2011a) used a radiative-conductive model inherited from Strobel et al. (1996) (and including convection in a putative troposphere in an updated version by Zalucha et al., 2011b) to fit observed occultation light curves from 1988, 2002, 2006, and 2008 in terms of $q_{CH4}$ and the surface radius. Perhaps not surprisingly, results were relatively "unstable", in the sense that large variations of $q_{CH4}$ were found (0.18-0.94 % in Zalucha et al. 2011a, and 0.2-0.7 % in Zalucha et al. 2011b) without showing a consistent evolution pattern. Still, this showed that even in the framework of these models, upper atmospheric temperatures of ~105 K can be satisfied by a large range of values of $q_{CH4}$.

There is ample observational evidence that dynamical redistribution of solar energy occurs in Pluto's atmosphere. Most stellar occultations observed since 1988 were multi-site events, with the different chords probing a variety of latitudes and morning/evening conditions. Once again, atmosphere temperatures show rather small fluctuations (< 10 K) from site to site (Millis et al. 1993, Young et al. 2008). In particular, although systematic studies in this direction are in progress (Zangari et al. 2012), no convincing correlation between upper atmosphere temperature and presumably relevant parameters (i.e. instantaneous insolation, surface albedo…) at occultation location has so far been evidenced. A striking example was the 2006 June 12 occultation (Young et al. 2008), for which the emersion temperature at 53°S, i.e. at the edge of the polar night, was indistinguishable from (and even nominally higher than) that at immersion, probing 30°N (106.4±4.6 K vs 100±4.2 K). All this indicates that dynamical redistribution of heat operates on timescales much shorter than radiative equilibrium. This observational fact is supported by theoretical calculations. Strobel et al. (1996) estimate radiative timescales to be of order 10-15 terrestrial years (still much shorter than a Pluto year), while for meridional winds of order 2 m/s (Zalucha and Michaels, 2013; Vangvichith 2013), dynamical timescales are of order of 10 terrestrial days only. In this situation, atmospheric dynamics ensures minimal horizontal variations in density and temperature (except in the planetary boundary layer). This makes globally averaged 1D radiative-conductive models adequate to calculate the middle atmosphere thermal structure.

In Paper I, a limit was put on the depth of Pluto's atmosphere (maximum pressure in 2008: 24 μbar; minimum Pluto radius: 1169 – 1172 km) based on the argument that too deep atmospheres are too cold to match the relative intensities of the various J-multiplets of the $2\nu_3$ band. Here, by examining the spectra in the regions of additional weak absorption, a more stringent limit on the radius (R > 1180 km) can be derived. This "large" radius is consistent with a value (R = 1173 ± 23 km) based on a re-assessement of mutual event data by Tholen and Buie (1997), and with more recent estimates (R = 1180 (+20/-10) km from Zalucha et al. (2011a), 1173 (+20/-10) km from Zalucha et al. 2011b).

The *New Horizons* spacecraft, scheduled to visit the Pluto system in 2015 with a closest fly-by on July 14, will measure among many other things Pluto's radius, surface pressure and methane abundance. For the sake (and amusement) of making predictions, we "guess" here a radius of 1184 km, based on our considerations on its likely lower (1180 km) and upper (1188 km) limits. For our "2012" profile, the pressure for this radius is 15 μbar. Accounting for further atmospheric expansion at the ~10 % level, the predicted surface pressure for 2015 is then 17 μbar. For $CH_4$, we rescale the best fit mean column density for 2012 (see Table 2) to a 1184 km radius, then multiply it by a factor 1.5 to account for its predicted evolution over 2012-2015 by the GCM (Fig. 6a). This gives a predicted 1.0 cm-am column and 0.44 % mixing ratio at the time of *New Horizons* encounter.

## 6. Summary

We have reported on repeated high-resolution (R ~ 60,000) observations of Pluto with VLT/CRIRES (2 spectra in 2008 and 5 spectra in 2012), acquired to constrain the spatial and vertical distribution of methane in Pluto's atmosphere and to search for mid-term (4 year) variability. A sensitivity study of the results (methane mixing ratio and column density) to model assumptions (temperature structure, surface pressure, Pluto radius) is performed. We reach the following conclusions:

- No variation of the $CH_4$ atmospheric content with Pluto rotational phase is detected in excess of 20 %.
- $CH_4$ column densities show at most marginal variations between 2008 and 2012, with a best guess estimate of a ~20 % decrease over this time frame. Stellar occultations indicate that Pluto's surface pressure has continued to increase over this period, implying a decrease of the methane mixing ratio. Our reassessment of the Young et al. (1997) observations gives further tentative evidence for a general decline of the methane content over 1992-2012.
- There is no evidence for an altitude-varying methane distribution. Rather, the methane mixing ratio is roughly uniform in at least the first 22-27 km of the atmosphere, and high concentrations of low-temperature methane near the surface can be ruled out. Our results are also best consistent with a relatively large (> 1180 km) Pluto radius.
- Comparison with predictions from a recently developed global climate model GCM of the methane cycle (Forget and Vangvichith 2013) indicates that these features are best explained if the source of methane occurs in regional-scale $CH_4$ ice deposits, including both low latitudes and high Northern latitudes, evidence for which is present from the rotational and secular evolution of the near-IR features due to $CH_4$ ice.

- Based on the above constraints, the observed evolution of surface pressure, and GCM predictions for the seasonal evolution of atmospheric methane, our “best guess” predictions for the *New Horizons* encounter in 2015 are: a 1184 km radius, a 17 μbar surface pressure, and a 0.44 % $CH_4$ mixing ratio (1.0 cm-am column), showing negligible variations with longitude.

# Figure Captions

Fig. 1. The 5 spectra observed in 2012 are shown in two intervals covering six R-lines, and compared to the 2 spectra from 2008. Data are shown in Pluto velocity scale. The position of the $CH_4$ lines from Pluto is shown by the dotted lines. They are flanked by their terrestrial counterparts (immediately to the left of the Pluto lines, except for the B spectrum where they are shifted to the right). The small vertical bars indicate the approximate 1-$\sigma$ error uncertainties.

Fig. 2. Thermal profiles (pressure-temperature-radius) used in the analysis. The squares (resp. circles) mark the 1188 km (resp. 1175 km) altitudes.

Fig. 3. Average spectrum (black histograms) observed in 2012, obtained by co-addition of the five individual spectra realigned in the Pluto velocity frame. Models are compared with fits obtained with the "2012" thermal profile and several values of the methane mixing ratio. Top row: surface radius = 1188 km. Bottom row: surface radius = 1175 km. This latter case over-predicts secondary methane absorption near 1643.32, 1645.77, and 1646.15 nm. From left to right: regions of the R7, R6 and R5 lines. The small vertical bars indicate the approximate 1-$\sigma$ uncertainties.

Fig. 4. Methane column densities (a) and mixing ratios (b) as a function of Pluto East longitude, for the two observing years, retrieved by assuming a surface radius of 1188 km and the respective thermal profiles for the two years, with surface pressures of 9.4 $\mu$bar in 2008 and 12.3 $\mu$bar in 2012. Data from 2012 are shown in green (and their average in blue). Data from 2008 are shown in red. These methane columns and mixing ratios can be rescaled to other assumptions for Pluto radius and surface pressure as indicated in the text. These methane results are shown in the context of other relevant indicators (arbitrary scale). Solid lines: Pluto's optical light curve in 2002-2003 (Buie et al. 2010a). Dashed-lines: depth of $CH_4$ ice strong (2.32 $\mu$m) band. Dashed-dotted lines: depth of $CH_4$ ice weak (0.97 $\mu$m) band. Dotted lines: degree of $CH_4$ dilution in $N_2$ ice, based on 1.72 $\mu$m band shifts (from Grundy et al. 2013). Background images are (a) map of Pluto (single scattering albedo) from HST observations in 2002-2003 (Buie et al. 2010b) (b) the Lellouch et al. (2000) distribution of Pluto units (white = $N_2$ ice, grey = $CH_4$ ice, black = tholins).

Fig. 5. Three spectral intervals of the August, 1 2008 spectrum, compared with models with different cut-off altitudes of the methane distribution (1210 km, 1225 km, no cut-off). The dark blue curve is the expected spectrum for the two-layer distribution of Cook et al. (2013), with 6.3 cm-am of methane at 40 K and 0.4 cm-am at 100 K. See text for details. The R1, R2, R3 and R5 lines of the $2\nu_3$ band are indicated. The small vertical bars indicate the approximate 1-$\sigma$ uncertainties.

Fig. 6. Rotational and temporal variations of the methane content predicted by the Pluto GCM of Forget and Vangvitchith (2013). a) Apparent variation of the mean methane mixing ratio as seen from Earth as a function of longitude. A very weak (+10 %) maximum is present near L =0. b) Time evolution of the global column density. Black line: nominal model, including

high-latitude methane ice. Red line: model, in which no $CH_4$ ice is present northward of 60°N. A 3 g $m^{-2}$ column is equivalent to 0.42 cm-am. Figures taken from Forget and Vangvitchith (2013).

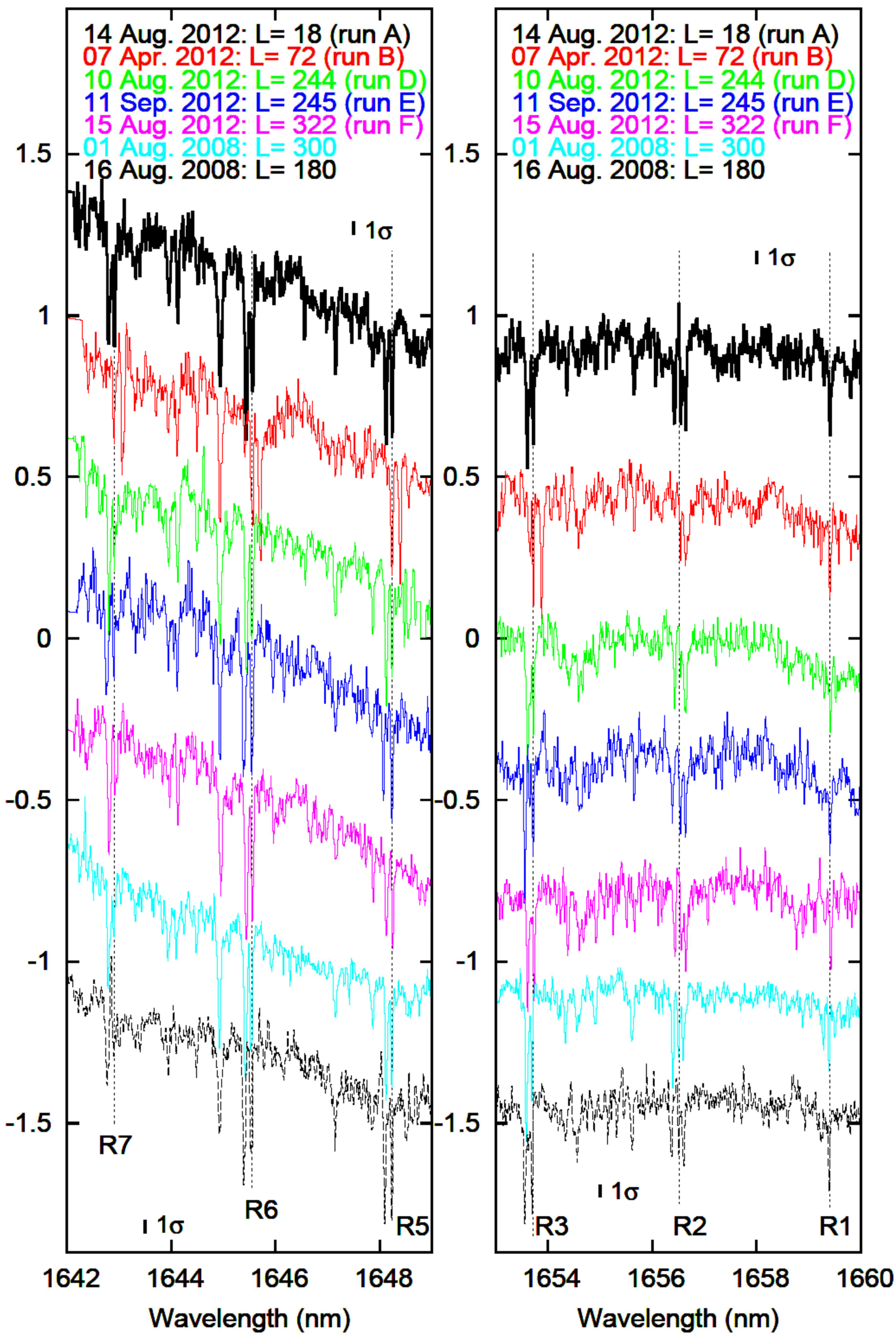
14 Aug. 2012: L= 18 (run A)
07 Apr. 2012: L= 72 (run B)
10 Aug. 2012: L= 244 (run D)
11 Sep. 2012: L= 245 (run E)
15 Aug. 2012: L= 322 (run F)
01 Aug. 2008: L= 300
16 Aug. 2008: L= 180
1σ
R7
R6
R5
R3
R2
R1
1.5
1
0.5
0
-0.5
-1
-1.5
1642
1644
1646
1648
1654
1656
1658
1660
Wavelength (nm)
Wavelength (nm)

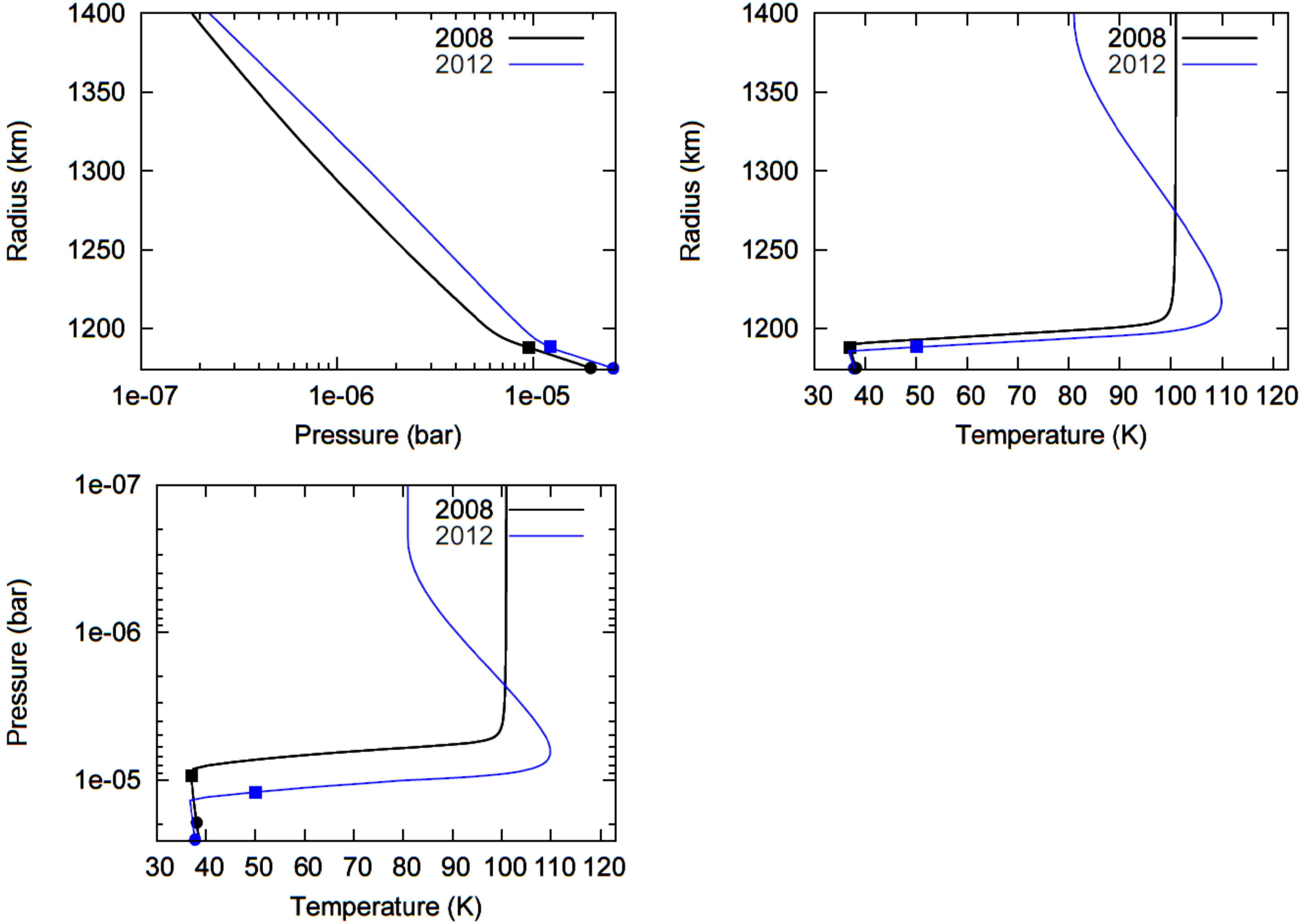

2008
2012
Radius (km)
Pressure (bar)
Temperature (K)
1400
1350
1300
1250
1200
1e-07
1e-06
1e-05
30 40 50 60 70 80 90 100 110 120

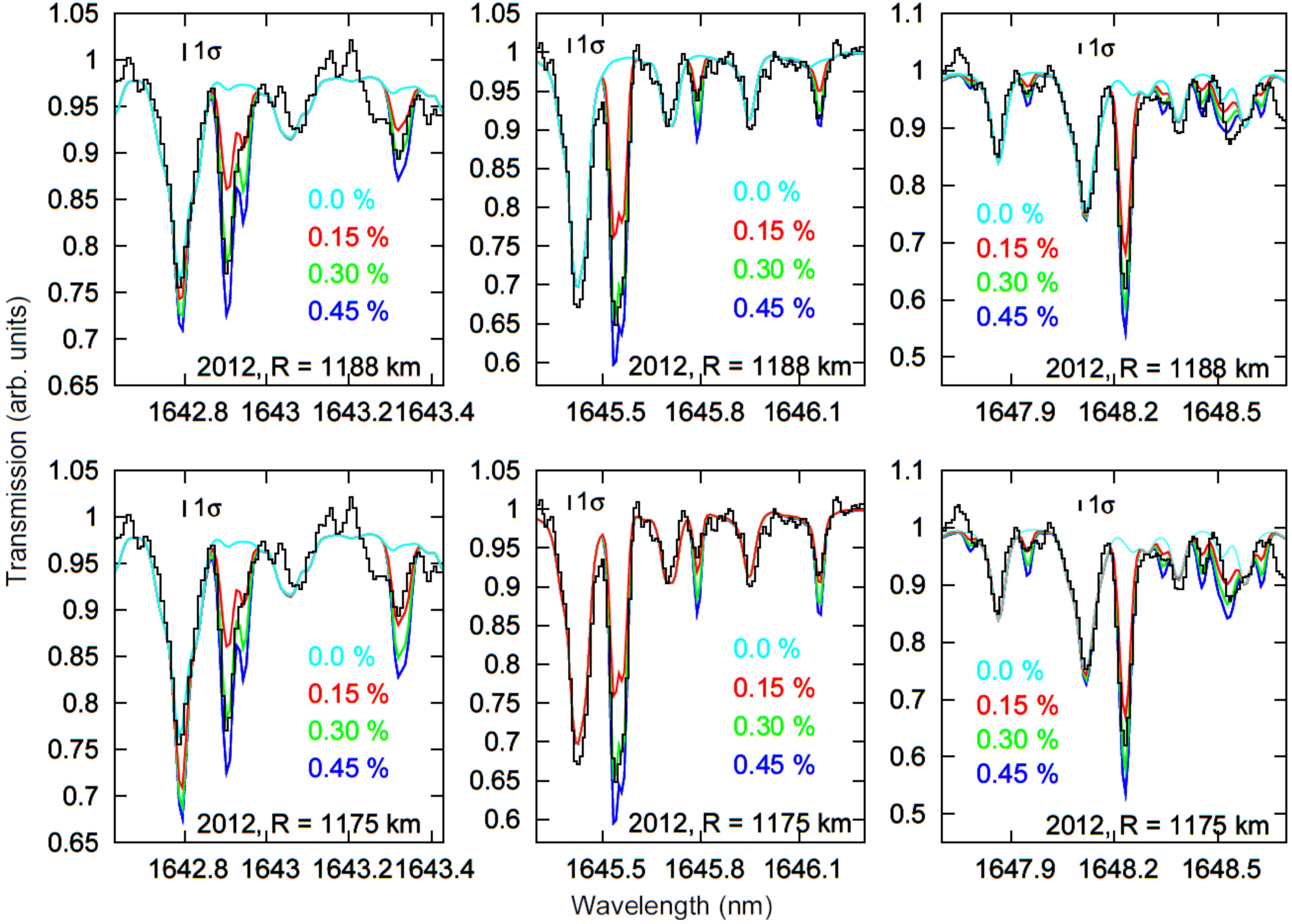

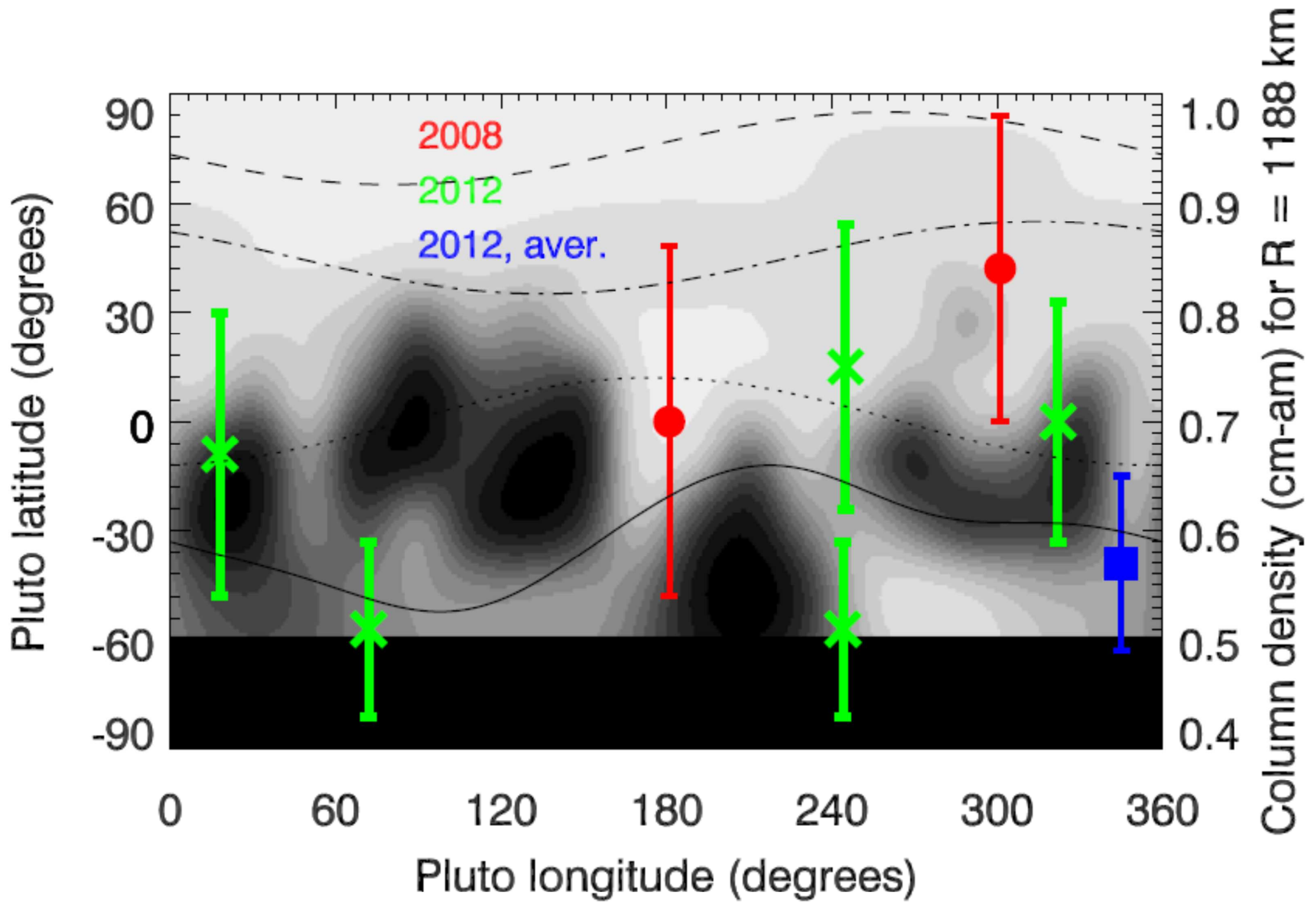

2008
2012
2012, aver.
Pluto latitude (degrees)
Pluto longitude (degrees)
Column density (cm-am) for R = 1188 km
90
60
30
0
-30
-60
-90
0
60
120
180
240
300
360
1.0
0.9
0.8
0.7
0.6
0.5
0.4

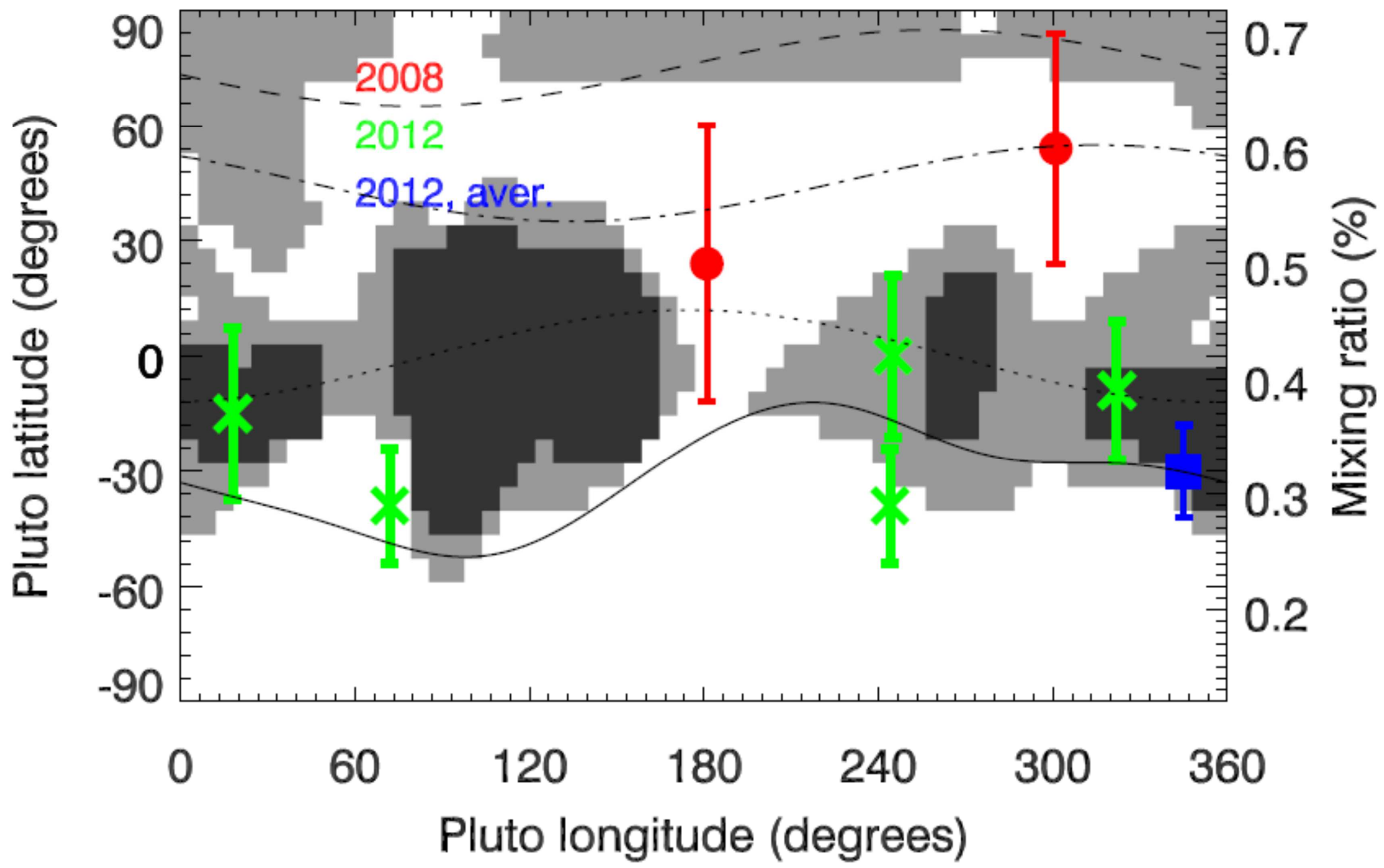

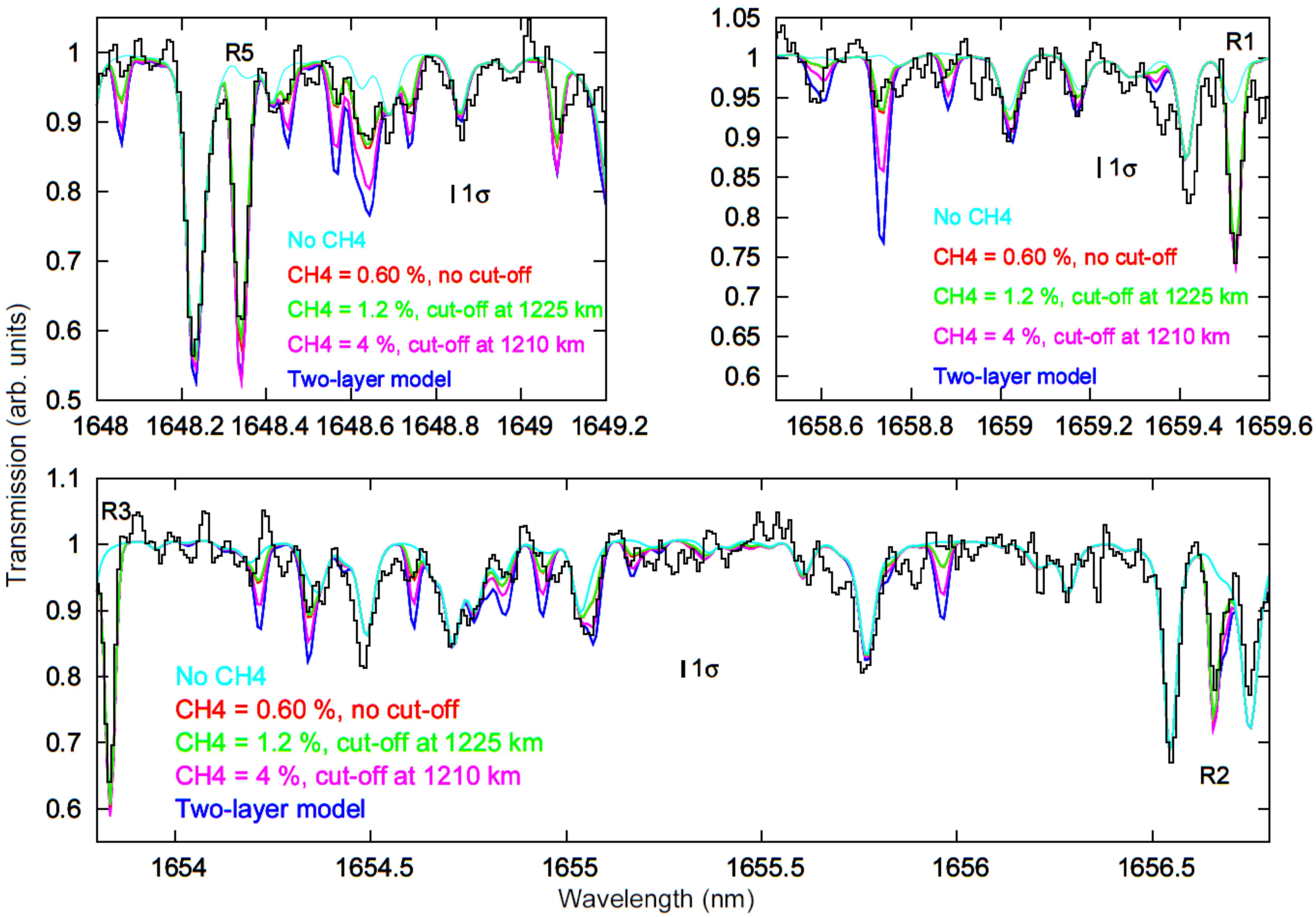

Transmission (arb. units)
Wavelength (nm)
R5
R1
R3
R2
I 1σ
No CH4
CH4 = 0.60 %, no cut-off
CH4 = 1.2 %, cut-off at 1225 km
CH4 = 4 %, cut-off at 1210 km
Two-layer model

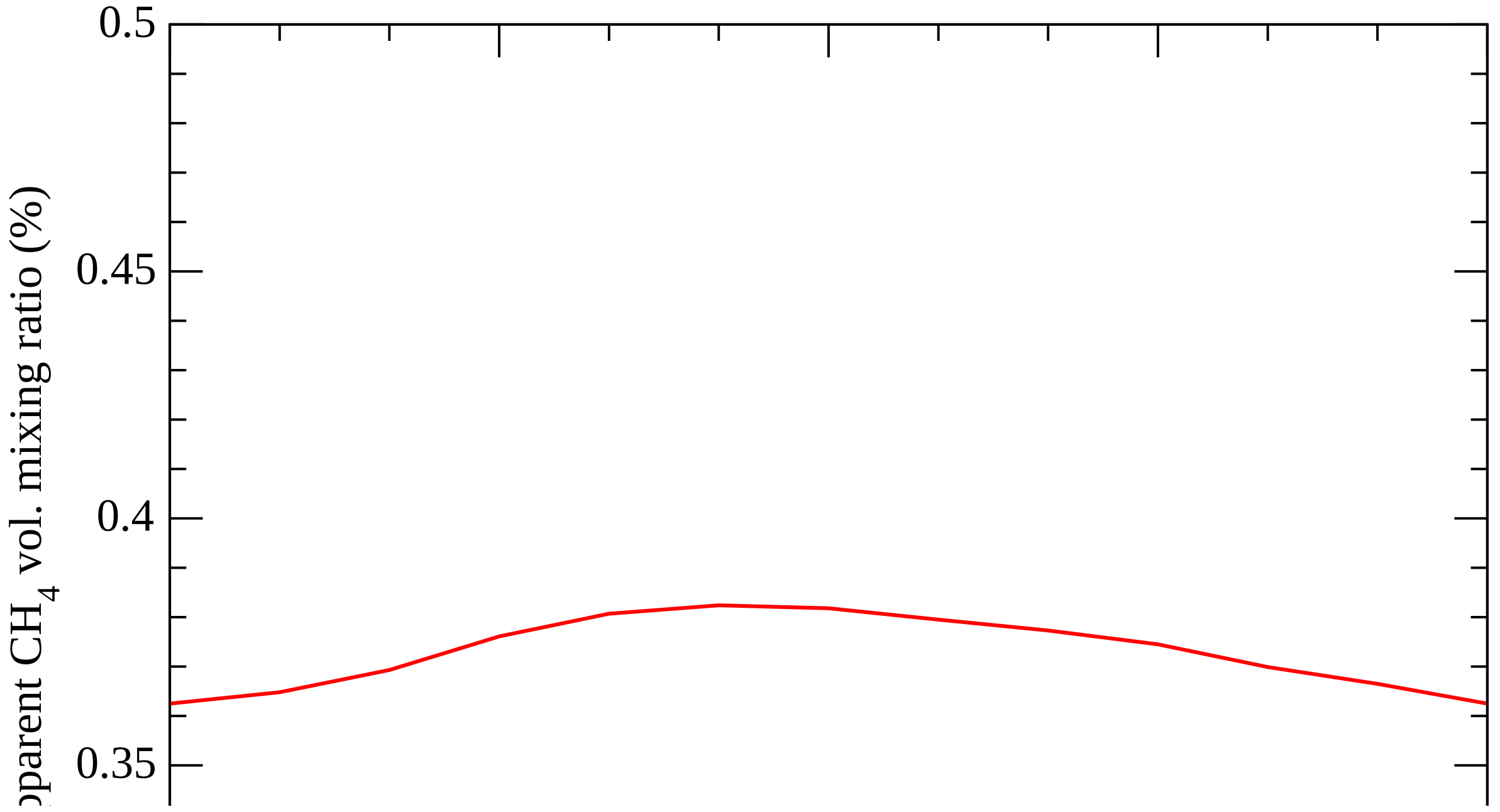

0.5
0.45
0.4
0.35
pparent $CH_4$ vol. mixing ratio (%)

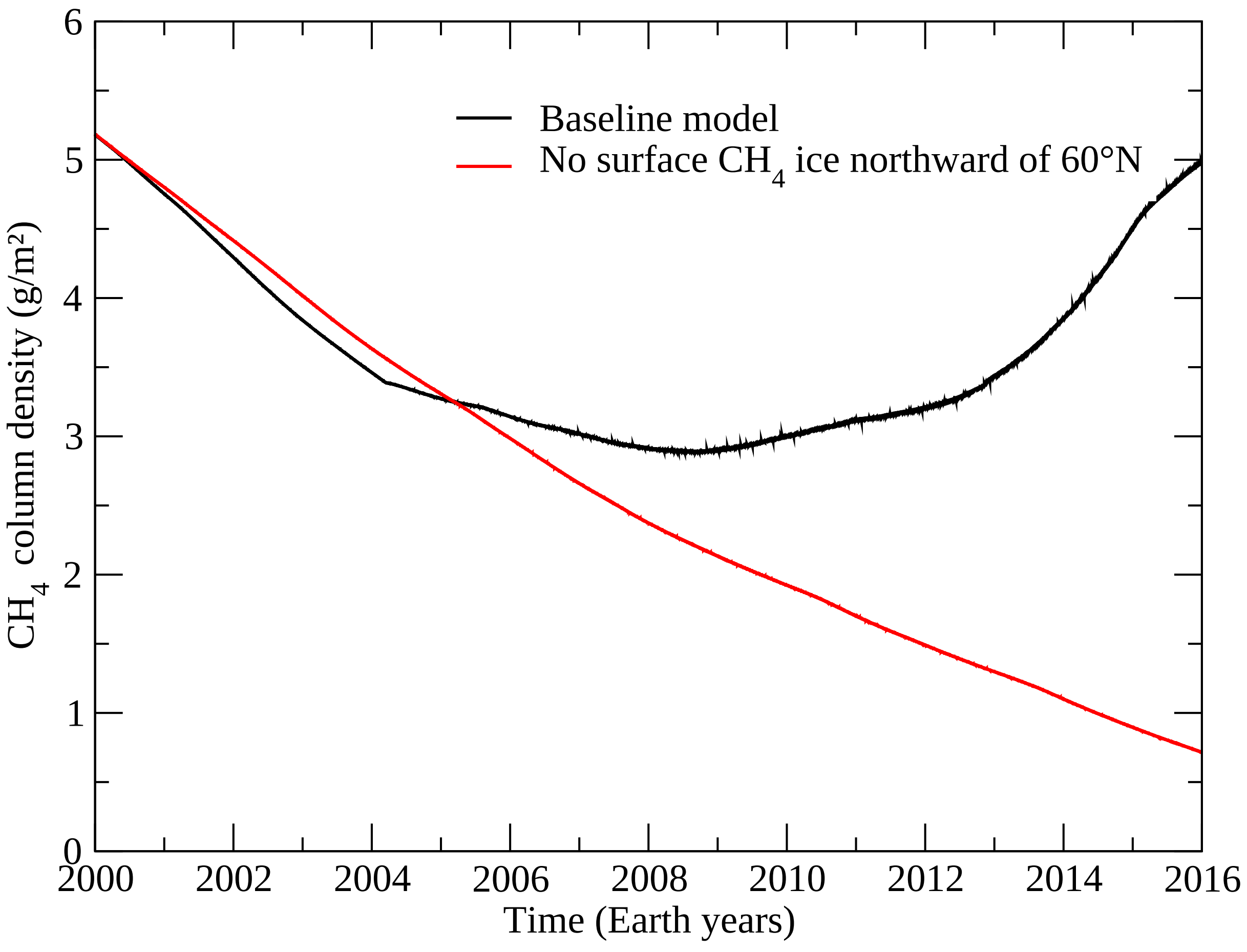